\documentclass{iaus}
\usepackage{graphicx}

\title[Gyrochronology for main sequence field stars] 
{Gyrochronology and its usage \\for main sequence field star ages}

\author[Sydney A. Barnes]   
{Sydney A. Barnes}

\affiliation{Lowell Observatory, 
\\1400 W. Mars Hill Road, Flagstaff, AZ 86001, USA
\\ email: {\tt barnes@lowell.edu} }

\pubyear{2008}
\volume{258}  
\pagerange{x01-x10}
\jname{Ages of Stars}
\editors{E.E. Mamajek, D.R. Soderblom \& J.A. Valenti, eds.}
\begin{document}

\maketitle

\begin{abstract}
The construction of all age indicators consists of certain basic steps which 
lead to the identification of the properties desirable for stellar age 
indicators. Prior age indicators for main sequence field stars possess
only some of these properties. The measured rotation periods of cool stars 
are particularly useful in this respect because they have well-defined 
dependencies that allow stellar ages to be determined with $\sim$20\% errors. 
This method, called gyrochronology, is explained informally in this talk, 
shown to have the desired properties, compared to prior methods, and used to 
derive ages for samples of main sequence field stars.

\end{abstract}

\firstsection 
\section{Motivations}
Things in our world come into being, exist in time, and eventually cease to be.
The properties of these things usually change over time, so that specifying 
the age of something, whether it be a tree, a human being, or a star, 
immediately gives us a good idea of what some of its other properties might be.
In galactic astronomy, the ages of individual stars assume a particular
importance, because they constitute the ticks of the abstract cosmic clock
that tells us how various astronomical phenomena change over time.

Sandage (1962, and earlier) first noted that the morphology of a cluster of 
stars in the Hertzsprung-Russell diagram might be used to derive its age. 
Demarque \& Larson (1964) improved it substantially, and named it the isochrone
method. Although venerable, this method is not very effective for main sequence 
stars because its principal variable, a star's luminosity, changes only slowly 
on the main sequence. Furthermore, for a field star, it also requires an 
excellent distance measurement, not easily accomplished. Consequently,
isochrone ages for main sequence field stars have errors approaching 
$\sim$100\%.

Of the prior distance-independent methods, the most consistent relies on the 
declining chromospheric activity of a cool star (Wilson, 1963; Skumanich 1972; 
Noyes et al. 1984; Soderblom et al. 1991; Donahue 1998). However, chromospheric
emission varies with a star's rotation phase, activity cycle phase, and
possibly other variables, limiting the precision of such ages to $\sim$50\%.

All activity-related age indicators are ultimately related to the rotation
rate of a star. However, attempts to harness rotation to derive ages were 
hindered by the ambiguity of $v \sin i$ measurements, and what we now know to 
be a dynamo-related bimodality, and associated transition, in very young stars.
The $v \sin i$ ambiguity can be entirely circumvented by (precisely) measuring 
a star's (mass-dependent) rotation period instead, and the early bimodality can 
be identified, and related stars excised. 

This method, named gyrochronology (and parsed {\it gyros-chronos-logos}) 
allows the derivation of a significantly more precise age than previously
available, of a cool main sequence field star, from its measured color and 
rotation period. 
The period is typically determined from time-series measurements of the 
spot-related photometric modulation of starlight. This talk is an informal 
summary of this method, as detailed in Barnes (2007). Some results and 
terminology derive from the `CgI scenario' for stellar rotation presented in 
Barnes (2003).

\section{Background for the construction of  all age indicators}

Many of the issues in constructing age indicators are so obvious that they are
routinely ignored! Let us therefore proceed by first stepping back, and 
considering the main steps in the construction of any age indicator. 
One needs to:
\begin{enumerate}
\item {\bf Find an observable, $v$, that changes `well' with age;}
      (`Well' means that it works for single objects rather than for an ensemble, 
      and also has the properties listed in Table~1.)
\item {\bf Determine the ages of suitable calibrators independently;}
      (This means measuring both the variable, $v$, for the calibrating objects, 
      and the most trustworthy prior variables so that $v$ can be related to 
      earth rotations, pendulum swings, etc.)
\item {\bf Measure the functional form of the variable: $v = v (t, w, x, ....)$;}
      ($t$ represents the age, and $w$, $x$, ... additional dependencies.
      Variables with the fewest dependencies are the most desirable.)
\item {\bf Invert that functional form to find $t = t (v, w, x, ...)$;}
      (Analytic inversions provide insight, but numerical inversions are usually 
      necessary.)
\item {\bf Calculate the error: $\delta t = \delta t (t, v, w, x, ....)$;}
      (Although necessary, non-linearities and other complexities often make this 
      final step difficult.)
\end{enumerate}

\begin{table}[h]
  \begin{center}
  \caption{Characteristics of the three major age indicators for field stars}
  \label{tab1}
 {\scriptsize
  \begin{tabular}{|l|l|l|l|}\hline 
{\bf Property$\Downarrow$\hspace{1cm}Method$\Rightarrow$}     & {\bf Isochrone Age} & {\bf Chromospheric Age} & {\bf Gyrochronology} \\ \hline
Measurable easily? & ? (Distance reqd.)  & ? (Repetition reqd.)    & ? (Repetition reqd.) \\ \hline
Sensitive to age?  &  No (on MS)	 & Yes                     & Yes                  \\ \hline
Insensitive to other parameters? &  No	               & Yes       & Yes                  \\ \hline
Technique calibrable?            & Yes (Sun)           & ? (Sun?)  & Yes (Sun)            \\ \hline
Invertible easily?               & No                  & Yes       & Yes                  \\ \hline
Errors calculable/provided?      &  ? (Difficult)      & Yes?      & Yes                  \\ \hline
Coeval stars yield the same age? & No (Field binaries) & ?         & Yes                  \\ \hline
  \end{tabular}
  }
 \end{center}
\end{table}

The foregoing, and other practical considerations, suggest that the following 
properties are desirable for {\it stellar} age indicators.
\begin{enumerate}
\item {\bf Measurability for single stars: } 
      The indicator should be properly defined, measurable easily itself, and 
      preferably should not require many additional quantities to be measured,
      otherwise it cannot be used routinely.
\item {\bf Sensitivity to Age: } 
      The indicator should change substantially (and preferably regularly) 
      with age, otherwise the errors will be inherently large.
\item {\bf Insensitivity to other parameters: } 
      The indicator should have insensitive (or separable) dependencies on other
      parameters that affect the measured quantity,
      otherwise there is the potential for ambiguity.
      In particular, distance-independent methods are preferred.
\item {\bf Calibration: }
      The technique should be calibrable using an object (or set of objects)
      whose age(s) we know very well, otherwise systematic errors will be
      introduced.
\item {\bf Invertibility: } The functional dependence determined above should
      be properly invertible to yield the age as a function of the measured
      variables.
\item {\bf Error analysis: } 
      The errors on the age derived using the technique ought to be calculable,
      otherwise no confidence can be attached to the ages.
\item {\bf Test of coeval stars: }
      The technique should yield the same ages for stars expected to be coeval,
      otherwise the validity of the technique itself must be questioned.
\end{enumerate}
\vspace{0.25cm}

Table\,1 summarizes the extent to which these properties are satisfied for the 
three major field star age indicators now available. 

\section{Introduction to rotational ages}

Skumanich (1972) seems to have been the first to identify a relationship
between the rotation rate of a star and its age. He used the averaged
$v \sin i$ values of stars in selected open clusters, and that of the 
Sun, all of whose ages are known independently. It was not clear then
that such a relationship could be used in any more than a statistical 
sense, partly due to the inherent ambiguity in $v \sin i$ measurements.
Observations of $v \sin i$ values, and later, rotation periods in young 
open clusters revealed a wide dispersion in the rotation rates of coeval 
stars that discouraged the use of rotation as an age indicator.

A prescient attempt was made by Kawaler (1989) to use rotation to derive 
ages based on the Hyades rotation period sequence, but its reliance on 
various theoretically motivated assumptions, the poor fit to the warm 
Hyades stars and the rotational dispersion in young open clusters cast 
doubts on its viability.

However, the availability of large numbers of rotation periods in
open clusters allowed the resolution of this `dispersion' into distinct
rotational sequences, C \& I, in color-period diagrams, each with its 
own set of dependencies (Barnes 2003). This resolution shows that the 
(largely slower-rotating) I sequence does indeed spin down similar to 
Skumanich's initial suggestion, but the (largely faster-rotating) 
C sequence does not. However, C sequence stars change into I sequence 
stars within a couple of 100\,Myr, so that all older cool stars must be 
of the I type, and spin down predictably. Furthermore, the spindown is 
convergent, in the sense that initial variations become increasingly 
unimportant with the passage of time.

These facts allow one to identify the principal dependencies of stellar
rotation, which turn out to be stellar color/mass and age, and to identify
empirically the tight relationship between them. This relationship must
be true for all cool stars on the main sequence. Therefore, measuring
a field star's color/mass and rotation period at once allow the age to
be determined. 

Furthermore, this method of determining the age is such that most of
the properties considered desirable for an age indicator, as listed
above, can be shown to be satisfied. Therefore, it seems appropriate 
to name the method `gyrochronology.'

\section{Color-period diagrams}

\begin{figure}[t]
\begin{center}
 \includegraphics[width=5.4in]{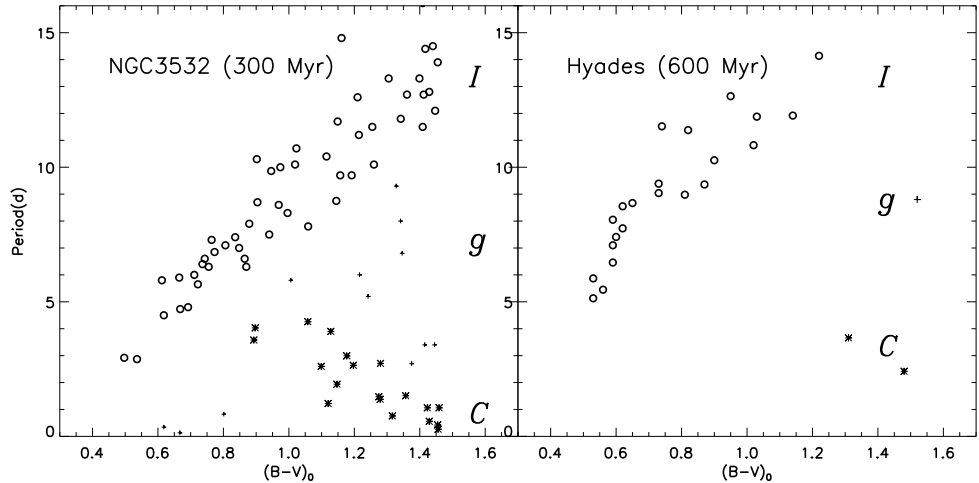} 
 \caption{Color-period diagrams of the 300\,Myr-old NGC3532 (Barnes, 1998)
and the 600\,Myr-old Hyades open clusters (primarily Radick et al. 1987). 
We see that the C-type stars in NGC\,3532 (asterisks) have changed into 
I-type stars (circles) by Hyades age.}
   \label{fig:figure1}
\end{center}
\end{figure}

Gyrochronology is ultimately based on color-period diagrams of open 
clusters, such as those shown in Fig.~1. 
(More such diagrams can be found in the papers by Meibom and Irwin in these
proceedings.)
The older ($\sim$600~Myr-old) Hyades cluster shows a distinct diagonal 
sequence, called I, of faster-rotating warmer stars and slower-rotating 
cooler stars, marked with circles.
The way to understand the color-period diagram of the younger 
($\sim$300~Myr-old) cluster NGC\,3532 is to realize that 
{\it not only is this I~sequence also present in this cluster, but 
another sequence, C, of faster-rotating stars}, marked with asterisks. 
[The sequences are striking in the richer M\,35 cluster (Meibom et al. 2008).]
Comparing the two color-period diagrams tells us that almost all the 
C-type stars change into I-type stars by Hyades age. 
The stars in the rotational gap, g, between the two sequences can now
be interpreted as stars in transition from the C- to the I-sequence.

These color-period diagrams also suggest that 
{\it the color/mass dependence of the I~sequence is the same for both clusters}.
This implies that the rotation period, $P$, of a star on this
I sequence is expressible as the separable product of this mass 
dependence and of other variables, of which we might guess that the
most important is the age, $t$, because stars spin down over time.
Thus, we write $P = f(B-V).g(t)$.

\section{The dependencies of I sequence stars}

What might the age dependence, $g(t)$, be? A very good guess would simply
be $g(t)= \sqrt{t}$, in agreement with the original suggestion by 
Skumanich (1972). Indeed, when the rotation periods, $P$, of stars in
all measured open clusters are divided by $g(t)=\sqrt{t}$, the I~sequences
are brought into coincidence, as shown in Fig.~2, from Barnes (2007).
(These early data include binaries, possibly aliased periods and other 
pathologies, hence the scatter.)

Fig.~3 shows a similar coincidence for field stars, the single unevolved set 
of Mt.\,Wilson stars from Baliunas et al. (1996). We have used individual 
chromospheric ages calculated using the prescription of Donahue (1998).
It is obvious that the C~sequence stars in the younger open clusters have
all changed into I~sequence stars in the older Mt.~Wilson sample. 
Furthermore, by guessing the age dependence, $g(t)$, using Skumanich (1972),
the mass dependence of the I~sequence has been made manifest.


\begin{figure}[t]
\begin{minipage}[b]{0.48\linewidth}
\centering
\includegraphics[width=2.3in]{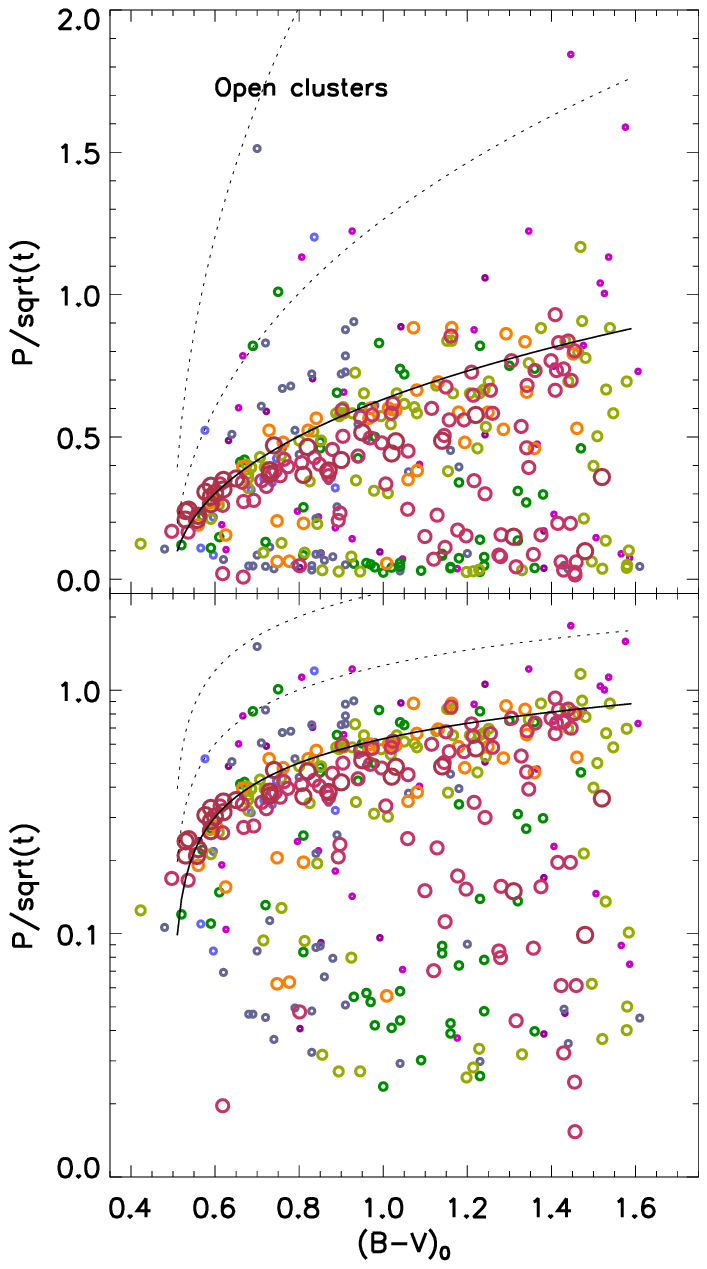}
\caption{Rotation periods of cluster stars divided by the square roots
of the cluster ages. Note the presence of both I~sequence stars near the
solid line, and C~sequence stars below. (Figure from Barnes, 2007)}
\label{fig:figure2}
\end{minipage}
\hspace{0.25cm}
\begin{minipage}[b]{0.48\linewidth}
\centering
\includegraphics[width=2.3in]{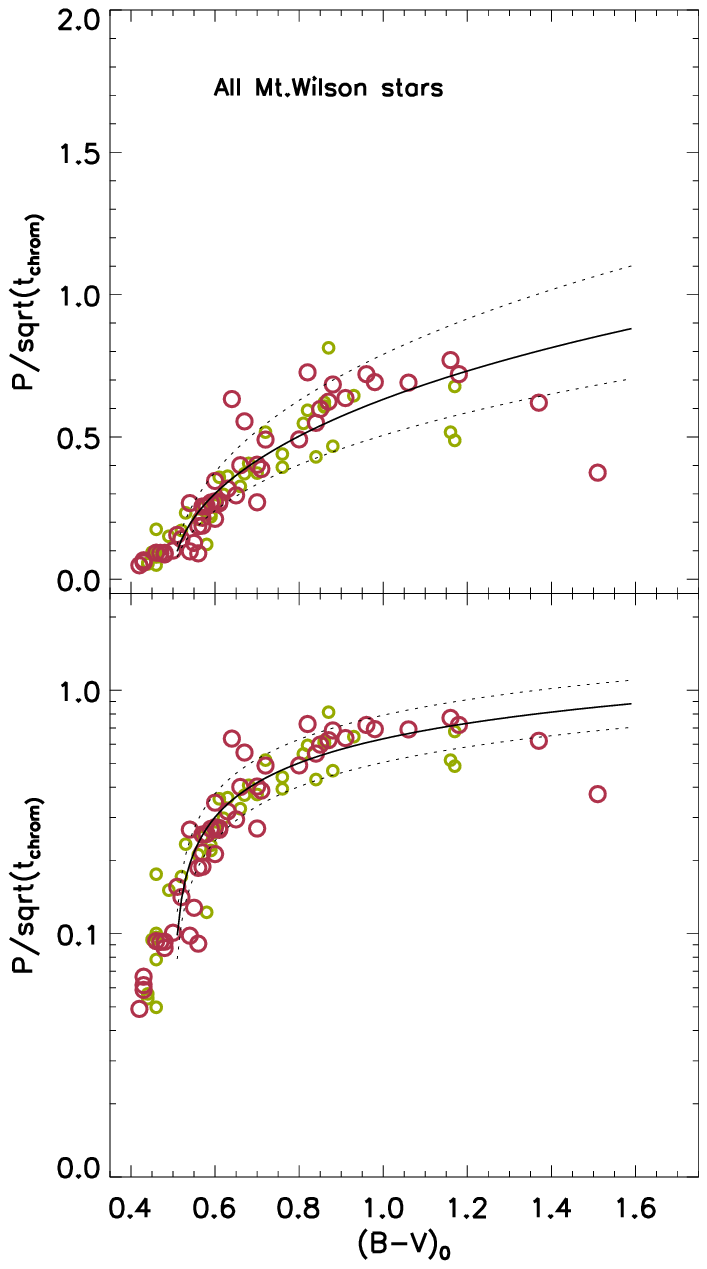}
\caption{Rotation periods of single, main sequence Mt.\,Wilson stars,
divided by the square roots of their chromospheric ages. Note that only
I~sequence stars are present. (Figure from Barnes, 2007)}
\label{fig:figure3}
\end{minipage}
\end{figure}

Indeed, one can fit this dependence using a function of the form
\begin{equation}
f(B-V) = a (B-V-c)^b \hspace{0.5cm}{\rm giving} \hspace{0.5cm} 
     a = 0.773 \pm 0.011, b = 0.601 \pm 0.024.
\end{equation}
The translational term, $c$, was simply equated to $0.4$ in 
Barnes (2007) and to $0.5$ in Barnes (2003).
A subsequent fit by Meibom et al. (2008), using both a large sample of 
rotation periods in the open cluster M\,35, and spectroscopic 
membership information, gives
\begin{equation}
a = 0.770 \pm 0.014, b = 0.553 \pm 0.052, c = 0.472 \pm 0.027.
\end{equation}
The point is that {\it regardless of the exact functional form chosen, 
a 2-3 parameter fit will suffice, and those parameters will be 
determined with small errors}.

One final step of the construction remains. Having specified $f(B-V)$, we
now return to $g(t)$. 
It is reasonable to seek a power law dependence: $g(t) = t^n$.
This allows us to calibrate the method using the Sun by ensuring 
that the above mass dependence gives the Solar rotation period at Solar age.
This calibration gives $n=0.519 \pm 0.007$.

Thus, the age of a star (in Myr) is simply given by inverting $P=f(B-V).g(t)$
to get
\begin{equation}
{\rm log}(t_{\rm gyro}) = \frac{1}{n} 
            \{ {\rm log}\,P - {\rm log}\,a - b \times {\rm log}\,(B-V-c) \}
\end{equation}
where the constants $a, b, c, n$ are as specified above, and base 10
logarithms are used.

\section{Age error analysis}

A virtue of the above formulation is that the age error can be simply 
calculated, and the various contributing error terms seen in perspective. 
The expression for the fractional age error, as calculated in Barnes (2007), 
is:
\begin{equation}
\frac{\delta t}{t} = 2\% \times 
                     \sqrt{ 3 + \frac{1}{2} ({\rm ln}\,t)^2  + 2 P^{0.6} 
                              + (\frac{0.6}{x})^2 + (2.4 {\rm ln}\,x)^2
                           }
\end{equation}
where $x = B-V-0.4$. 
For 1\,Gyr-old stars of spectral types late\,F, early\,G, mid\,K and early\,M 
respectively, we get
\begin{equation}
{
\frac{\delta t}{t} = 2\% \times 
\cases{
\sqrt{26.9 + 6.4 + 66.5}  & when $B-V=0.5$  ($P= 7d$);\cr
\sqrt{26.9 + 8.9 + 16.9}   & when $B-V=0.65$ ($P=12d$);\cr
\sqrt{26.9 + 12.1 + 2.5}  & when $B-V=1.0$  ($P=20d$);\cr
\sqrt{26.9 + 15.4 + 0.35}  & when $B-V=1.5$  ($P=30d$).\cr
      }
}
\end{equation}
which shows the relative contributions of the period and color errors (second
and third terms, respectively). Color errors and differential rotation are the
significant contributors for bluer and redder stars respectively.

The expression above evaluates to fractional age errors of 
13-20$\%$ for 1\,Gyr-old early\,M-late\,F stars, suggesting that relatively
precise ages may indeed be derived for field stars, provided that the
observable inputs, color and rotation period, are measured well.

\section{Application to field star samples}

Expressions (5.3) above and (6.1), for the gyro age and its error, 
respectively, are true for all I-type main sequence late\,F-early\,M stars, 
whether in clusters or in the field. 
We can therefore apply them to field star samples with measured rotation 
periods to derive ages where none were available before.
The field star sample of Strassmeier et al. (2000) is an example.

\begin{figure}[h]
\begin{minipage}[b]{0.48\linewidth}
\centering
\includegraphics[width=2.3in]{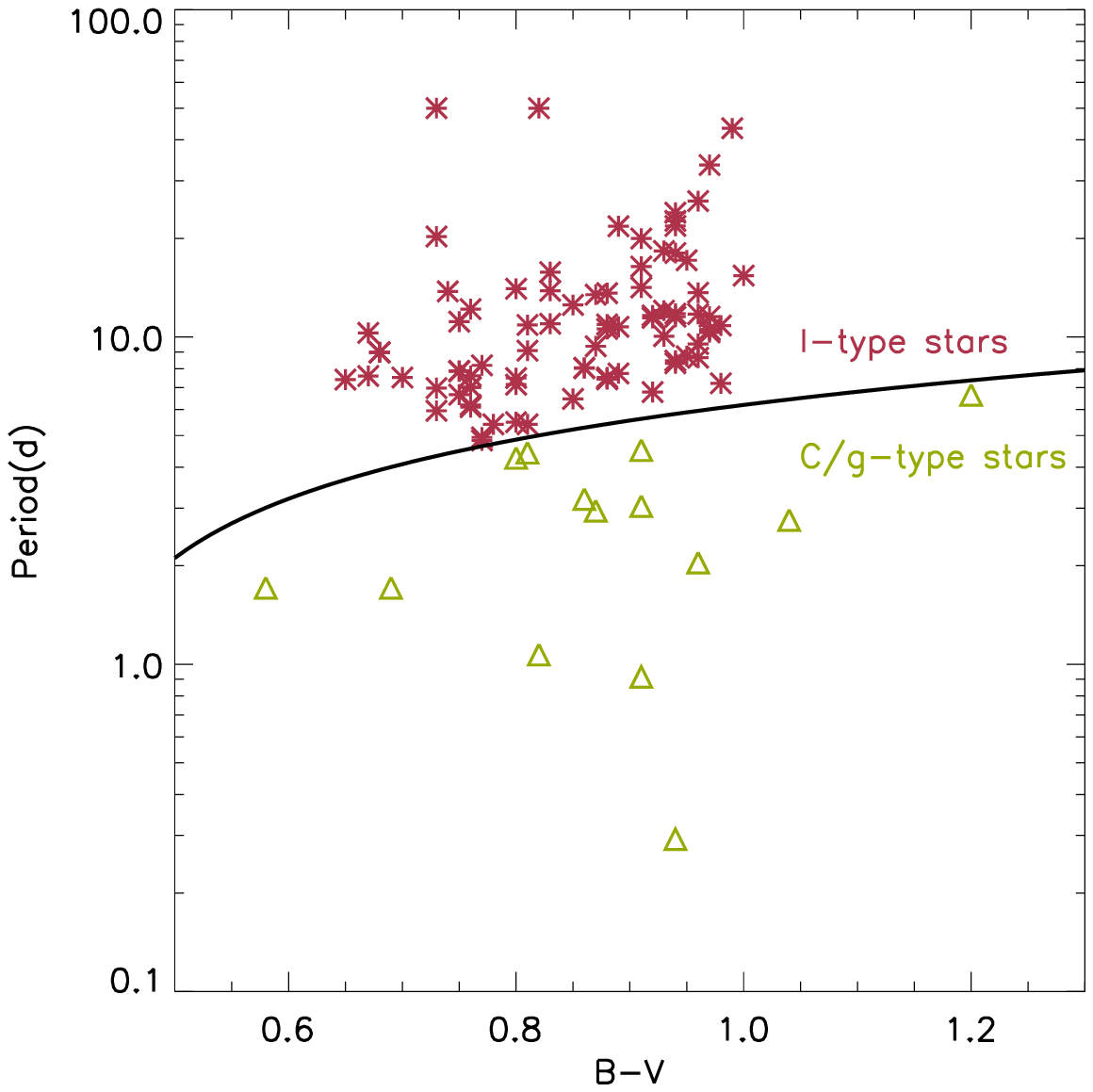}
\caption{Color-period diagram for the Strassmeier et al. (2000) stars, 
showing the 100\,Myr-old (gyro) isochrone used to discard possible C/g-type 
stars. Only the I-type stars above are retained for gyrochronology.
(Figure from Barnes, 2007)}
\label{fig:figure4}
\end{minipage}
\hspace{0.25cm}
\begin{minipage}[b]{0.48\linewidth}
\centering
\includegraphics[width=2.3in]{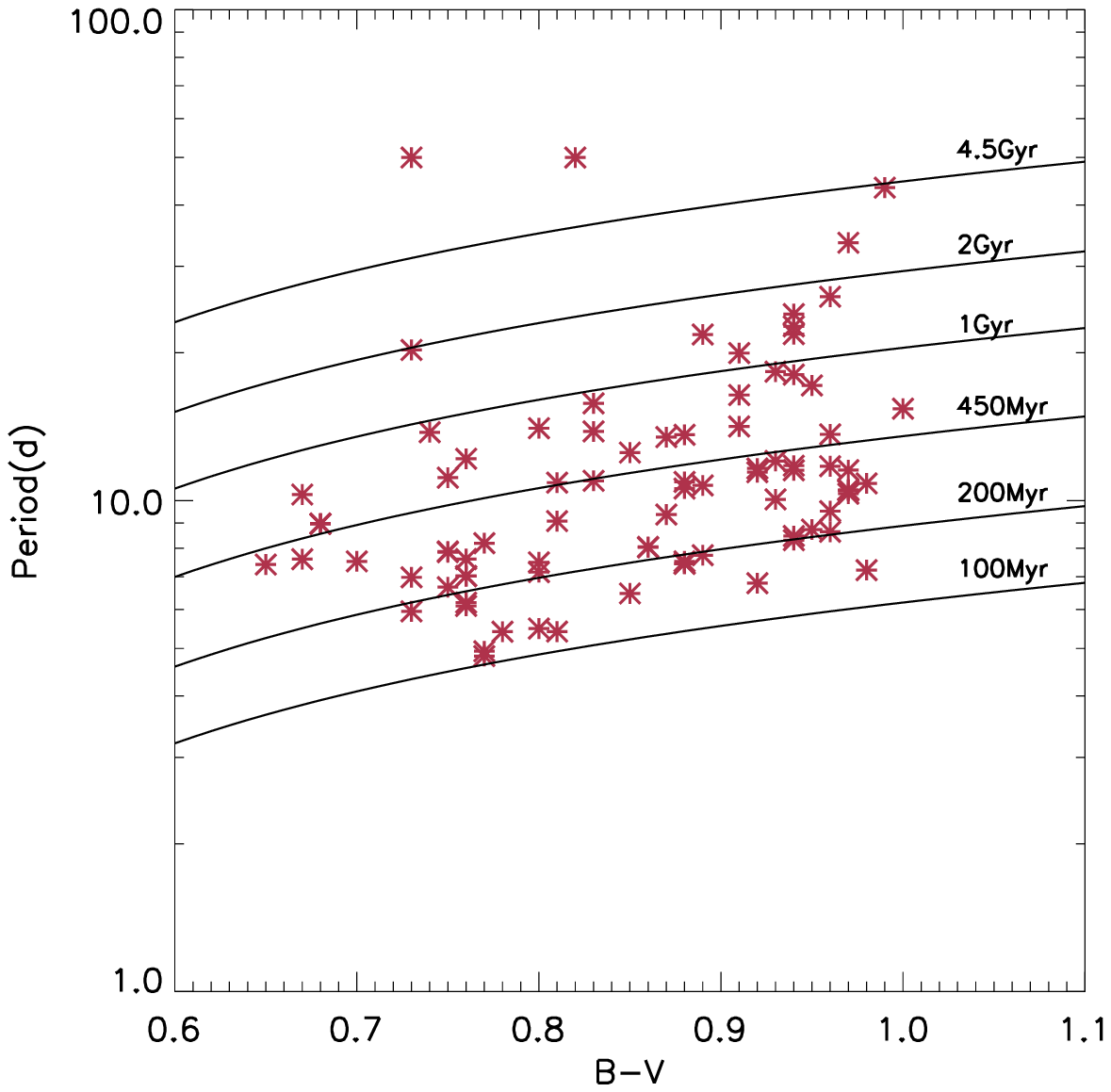}
\caption{Rotational isochrones are shown for ages ranging from 100\,Myr
to 4.5\,Gyr for the I-type stars in the Strassmeier et al. (2000) sample.
Note the relative youth of the sample, in keeping with its active pedigree.
(Figure from Barnes, 2007)}
\label{fig:figure5}
\end{minipage}
\end{figure}

Fig.\,4 displays the color-period diagram for this sample, and the 100\,Myr
(gyro) isochrone used to choose only the I-type stars for age analysis.
Fig.\,5 displays isochrones spanning the age range of the sample, showing
its relative youth. Indeed, the median age is only 365\,Myr, in agreement with
the selection of the original sample by activity. 
The stars are individually tabulated in Barnes (2007), 
where the technique is also applied to the older sample of stars
(median age of 1.2\,Gyr) assembled by Pizzolato et al. (2003).
For both samples, activity indicators like $R_{HK}$ and $L_X/L_{bol}$ are found
to decline as expected with increasing gyro age.

\section{Comparison with chromospheric ages}

The best age indicator for nearby field stars over the past couple of
decades has been the decline of chromospheric emission with age.
The calibrations regularly used are those of Soderblom et al. (1991) and
Donahue (1998), but see Mamajek \& Hillenbrand (2008) and Mamajek's
article in these proceeding for a recalibration including gyrochronology.
It would therefore be appropriate to compare the new gyro ages with these 
older chromospheric ages.  
The best sample for this comparison is the Mt.\,Wilson sample of cool stars,
one studied intensively for decades for chromospheric activity, and for which
measured rotation periods are also available.

Fig.\,6 shows this comparison, the cross indicating representative errors.
The basic result to note is that there is reasonable agreement between the
two ages because the upper left and lower right corners are unoccupied.
A closer inspection shows that the chromospheric ages used here (Donahue 1998)
are somewhat longer than the gyro ages, as the dashed median line shows.
Dividing the stars into blue ($B-V < 0.6$), green ($0.6 > B-V > 0.8$),
and red ($B-V>0.8$), shows that the discrepancy relates mostly to the blue
F\,stars, whose lifetime of 5\,Gyr is indicated in the figure, as is the
age of the universe.

\begin{figure}[b]
\begin{center}
 \includegraphics[width=3.4in]{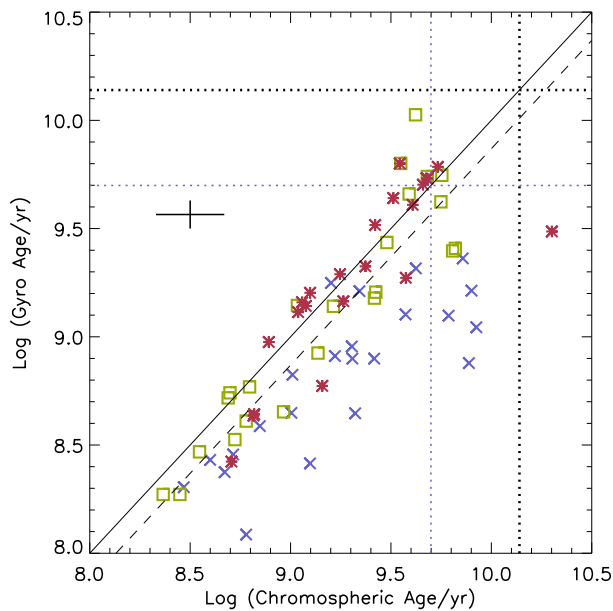} 
 \caption{Comparison between gyro- and chromospheric ages for the Mt.\,Wilson 
star sample. Two conclusions can be drawn: (1) The two types of ages are in 
rough agreement, but the chromospheric ages are somewhat larger, and (2) The 
discrepancy relates mostly to the blue F stars (crosses) whose 5\,Gyr lifetime is 
marked, rather than the green G stars (squares) or the red K stars (asterisks).
(Figure from Barnes, 2007)}
   \label{fig:figure6}
\end{center}
\end{figure}

\section{Comparison with isochrone ages}

An equivalent comparison of gyro- and isochrone ages demonstrates the difficulty 
of deriving isochrone ages for field stars.

The most modern and homogeneous field star isochrone ages available are those
for the SPOCS star sample of Takeda et al. (2007), who have undertaken a 
Bayesian age analysis, based on the method of Pont \& Eyer (1994), and a prior
uniform spectroscopic study of these stars by Valenti \& Fischer (2005).
The stars in common with those in Barnes (2007) are displayed in Fig.\,7.

Despite the Bayesian technique's admirable attempt to account for the asymmetric 
error distribution in color-magnitude diagrams, the isochrone ages are still on 
average a factor of $\sim$2.7 larger than the gyro ages.
Some upper and lower limits are included when they represent wide binary stars
with measured rotation periods. Related components are connected with dashed
lines. These only serve to underscore the difficulty of deriving isochrone ages
for non-cluster main sequence stars.

\begin{figure}[h]
\begin{center}
 \includegraphics[width=3.4in]{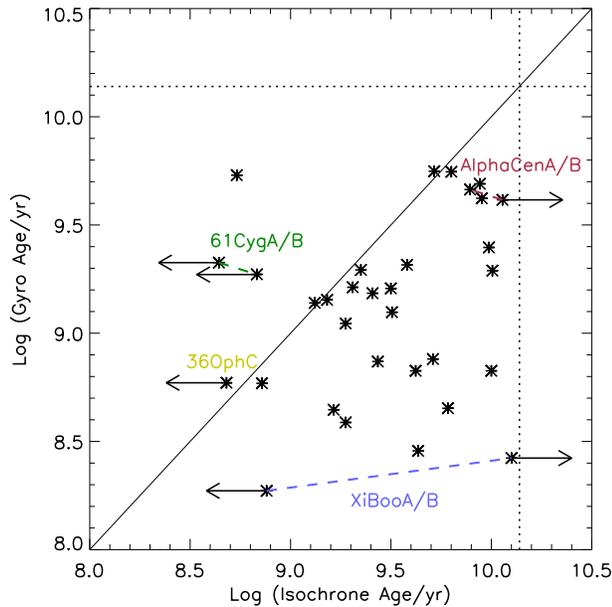} 
 \caption{Comparison between gyro- and Bayesian isochrone ages for stars in 
common with the Takeda et al. (2007) SPOCS sample. The isochrone ages are on
average $\sim$2.7 times the gyro ages. Upper- and lower limits are included if
they concern wide binaries with rotation periods. 
The components are connected by dashed lines. (Figure from Barnes, 2007)}
   \label{fig:figure7}
\end{center}
\end{figure}

\section{Ages for wide binaries}

Finally, we arrive at that very desirable property that an age determination 
method yield the same age for stars that we believe to be coeval.
Indeed, there are a handful of wide binaries where rotation periods for both
components have been measured. Thus, their ages may be determined independently.
(We use wide binaries to be sure that there has been no tidal or magnetic
hanky-panky between the components.)

The color-period diagram for the three available systems is shown in Fig.\,8,
along with the mean gyro isochrones and their errors for each pair, with
details in Table\,2 (from Barnes 2007). 
The ages for the components appear to be in agreement within the errors. 
The age of the 36\,Oph triple system is taken to be that of the presumably
non-interacting tertiary component.

\begin{figure}[t]
\begin{center}
 \includegraphics[width=3.4in]{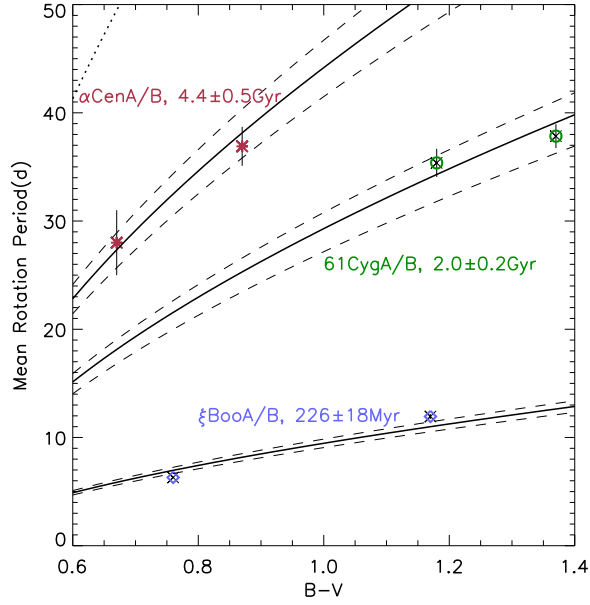} 
 \caption{The components of the wide binaries $\xi$ Boo, 61\,Cyg, and $\alpha$ Cen 
appear to give the same gyro ages. (Figure from Barnes, 2007)}
   \label{fig:figure8}
\end{center}
\end{figure}

\begin{table}[b]
  \begin{center}
  \caption{Ages for wide binary systems}
  \label{tab2}
 {\scriptsize
  \begin{tabular}{|r|r|l|r|r|r|r|}\hline 
{\bf System} & {\bf Star}     & {\bf $B-V$} & {\bf $\bar{P}_{rot}$} & {\bf $Age_{chromo}$} & {\bf $Age_{iso}$} & {\bf $Age_{gyro}$} \\ \hline
       &HD131156A & 0.76 & 6.31(0.05)  & 232\,Myr &$<$760\,Myr    & 187$\pm$21\,Myr\\
$\xi$ Boo&HD131156B & 1.17 & 11.94(0.22) & 508\,Myr &$>$12600\,Myr  & 265$\pm$28\,Myr\\
       &Mean       &      &           &          &         & {\bf 226$\pm$18\,Myr}\\
\hline
       &HD201091 &1.18  &35.37(1.3) & 2.36\,Gyr &$<$0.44\,Gyr  & 2.12$\pm$0.3\,Gyr\\
61 Cyg &HD201092 &1.37  &37.84(1.1) & 3.75\,Gyr &$<$0.68\,Gyr  & 1.87$\pm$0.3\,Gyr\\
       &Mean      &       &            &           &     & {\bf 2.0$\pm$0.2\,Gyr}\\
\hline
          &HD128620 & 0.67 & 28(3)   & 5.62\,Gyr   &7.84\,Gyr     & 4.6$\pm$0.8\,Gyr\\
$\alpha$ Cen&HD128621 & 0.87 & 36.9(1.8) & 4.24\,Gyr &$>$11.36\,Gyr & 4.1$\pm$0.7\,Gyr\\
          &Mean     &      &            &            &     & {\bf 4.4$\pm$0.5\,Gyr}\\
\hline \\
\hline
       &HD155886 & 0.85 & 20.69(0.4) & 1.1\,Gyr  & ......      & 1.42$\pm$0.19\,Gyr\\
36 Oph &HD155885 & 0.86 & 21.11(0.4) & 1.2\,Gyr  & ......      & 1.44$\pm$0.20\,Gyr\\
      &HD156026 & 1.16 & 18.0(1.0) & 1.4\,Gyr &$<$0.48\,Gyr &{\bf 0.59$\pm$0.07\,Gyr}\\
\hline
  \end{tabular}
  }
 \end{center}
\end{table}

\section{Conclusions}

In summary, we have constructed an improved method of determining the age of a
main sequence star from its measured rotation period, calibrated it using the
Sun, and shown that the associated errors are smaller than those from prior
methods.

The key steps of the construction are:
\begin{itemize}
\item All late\,F-early\,M stars become I-type rotators within a couple of 100\,Myr,
\item Their rotation periods, $P$, are describable as a product of two separable 
  functions, $f$, and $g$, of the $B-V$ color and age, $t$, respectively:
  $P(B-V, t) = f(B-V) . g(t)$,
\item $f(B-V)$ and $g(t)$ can be determined empirically with small errors,
\item $g(t)$ is such that initial variations become increasingly irrelevant with time,
\item The functional dependence is easily inverted to get $t = t (P, B-V)$, and
\item The age error, $\delta t = \delta t (t, P, B-V)$, is calculated.
\end{itemize}

The technique compares favorably with prior methods, which it complements, 
and passes some important tests. Precise time-series photometry is 
increasingly available from the ground and from space, making stellar 
rotation periods routinely measurable. 
Consequently, we recommend measuring rotation periods for appropriate
main sequence cool stars where precise ages are desired, and using 
gyrochronology to derive them.

\end{document}